\documentclass{article}
\usepackage{amssymb}
\usepackage{amsmath}

\input{tcilatex}

\begin{document}

\title{Quantum phenomena inside a black hole: quantization of the scalar
field iniside horizon in Schwarzschild spacetime}
\author{Pawel Gusin$^{1}$, Andrzej Radosz$^{1}$, Andy T. Augousti$^{2}$,
\and Janos Polonyi$^{3}$, Oleg B. Zaslavskii$^{4}$ and Romuald J. \'{S}%
ciborski$^{5}$ \\
$^{1}$Department of Quantum Technologies, \\
Wroclaw University of Science and Technology, Wroclaw, Poland\\
$^{2}$Faculty of Science, Engineering and Computing, \\
Kingston University London, London, UK\\
$^{3}$CNRS-IPHC, Strasbourg University, 23 r. du Loess BP28 67037, \\
Strasbourg Cedex 2, France\\
$^{4}$Department of Physics and Technology, Kharkov V.N. Karazin\\
National University, 4 Svoboda Square, Kharkov 61022, Ukrainey\\
$^{5}$Jaramogi Oginga Odinga University of Science and Technology in\\
Bondo (Kenya) }
\maketitle

\begin{abstract}
\textit{We discuss the problem of the quantization and dynamic evolution of
a scalar free field in the interior of a Schwarzschild black hole. A unitary
approach to the dynamics of the quantized field is proposed: a
time-dependent Hamiltonian governing the Heisenberg equations is derived. It
is found that the system is represented by a set of harmonic oscillators
coupled via terms corresponding to the creation and annihilation of pairs of
particles and that the symmetry properties of the spacetime, homogeneity and
isotropy are obeyed by the coupling terms in the Hamiltonian. It is shown
that Heisenberg equations for annihilation and creation operators are
transformed into ordinary differential equations for appropriate Bogolyubov
coefficients. Such a formulation leads to a general question concerning the
possibility of gravitationally driven instability, that is however excluded
in this case.}
\end{abstract}

\section{Introduction}

The horizon of a black hole (BH) may be regarded as a geometrical
singularity ("fake geometrical singularity"). Indeed, considering a
Schwarzschild BH in Schwarzschild coordinates one finds the metric tensor
exhibiting an on-horizon singularity that is absent in other,
singularity-free coordinate systems. There are a variety of singularity-free
coordinate systems in this case, e.g. Kruskal-Szekers,
Eddington-Finkelstein, Novikov and others [1-2]. Two interesting
observations might be made here. The first is to notice that the presence of
the event horizon is manifested both in coordinates revealing the horizon's
singularity as well as in the singularity-free systems. The second one is to
note the surprising similarities and/or analogies for phenomena taking place
outside and inside black holes. A rather well-known example of such a
property is the so-called BSW effect [3]. Two-particle collisions occuring
in the vicinity of the black hole's horizon may lead to a high-energy
outcome according to two scenarios [4-5]. These two scenarios turn out to be
the same in the exterior as well as in the interior of BH. A variety of
other aspects of the Exterior vs Interior (a)symmetry have been discussed in
Ref. [6].

It was shown by Doran et al. [7] that the interior of a Schwarzschild BH's,
which is a dynamically changing spacetime, may be regarded as a solution of
Einstein's equation. This interior spacetime, also called "T-sphere" (see
[8]) which is globally hyperbolic, gains then the status of a cosmological
model. Its 3D spatial-like section is a hypercylinder $\mathbf{R}^{1}\times
S^{2}$, expanding longitudinally, along the homogeneity direction $\mathbf{R}%
^{1},$ (see also [6-8]) and contracting transversally, perpendicularly to
this direction in the angular coordinates of the sphere $S^{2}$. However, as
shown in Ref. [7], such a process may be preceded by a process of expansion
of the sphere and collapsing of the cylinder to its base sphere of radius $%
r_{S}$. Such an expansion followed by a contraction constitutes the full
cycle for the cosmological model introduced in [7].

Various phenomena and processess have been considered both in the interior
of the Schwarzschild BH [8, 10-13] and in its extension [7] to which we will
hereafter refer to as the "T-model", an anisotropic cosmological model. In
particular the Yang-Mills and Higgs fields in the Kantowski-Sachs
anisotropic, cigar-like -- referred to above as a hypercylinder --
cosmological model were discussed in [14] (see also [15]). Canonical
quantization of the scalar field inside a Schwarzschild BH was presented by
Yajnik and Narayan [16], where a so-called tortoise coordinate was used, in
consequence leading to a Hamiltonian of diagonal form and, as claimed by the
authors, to "QFT set up by the freely falling observer". Other studies of
the quantum properties of scalar field were given for instance in Refs.
[17-18] and the investigations of the interior of the Schwarzschild BH were
presented in Refs. [19-20]. The most recent results have been given by
Almeida and Rodrigues in Ref. [21] where the quantization of the BH gravity
was discussed and by Giddings and Perkins in Ref. [22], in which the quantum
evolution of the Hawking state in Schwarzschild spacetime was investigated.

In this paper we will present a particular quantum aspect of the "T-model".
Namely the problem of dynamics, i.e. the temporal evolution of the quantized
scalar field in the case of such a cosmology will be introduced and briefly
discussed within a unitary approach. The Hamiltonian of the system,
represented by a set of harmonic oscillators, coupled via creation and
annihilation of pairs of particles, revealing interesting symmetry
properties, will be derived. The Heisenberg equations of motion for
appropriate annihilation and creation operators will be converted into
ordinary differential equations for Bogolyubov coefficients and will be
shown to reveal the possibility of an instability that is referred to as a
gravitationally driven instability.

The paper is organized as follows. In Sec. 2 we discuss the properties of
the Schwarzschild BH and a T-model is formulated. In Sec. 3 a scalar field
and its quantization are discussed. In Sec. 4. the Hamiltonian of the scalar
field is derived and a discussion is presented in the final section, Sec.5;
Appendix is devoted for a derivation of explicit form the temporal part of
(factorized) Klein-Gordon equation.

\section{"T-sphere" model - an anisotropic cosmological model}

The metric $g_{\mu \nu }$ for the exterior of the Schwarzchild black hole,
diagonal in the Schwarzschild coordinates $\left( t,r,\theta ,\varphi
\right) $, reveals the singularity on the horizon:%
\begin{equation}
ds^{2}=g_{t}\left( r\right) dt^{2}-g_{r}\left( r\right) dr^{2}-g_{2}\left(
r\right) d\Omega ^{2}.  \tag{2.1}
\end{equation}%
where 
\begin{equation}
g_{t}=1-\frac{2M}{r}=g_{r}^{-1}  \tag{2.2}
\end{equation}%
$g_{2}\left( r\right) =r^{2},$ and $d\Omega ^{2}$ denotes the metric on the
two-dimensional unit sphere $S^{2}$ with the coordinates $\left( \theta
,\phi \right) :$%
\begin{equation}
d\Omega ^{2}=d\theta ^{2}+\sin ^{2}\theta d\phi ^{2}.  \tag{2.3}
\end{equation}%
The geometrical singularity at the horizon, $r_{S}=2M$ may be removed by a
transformation to a singularity-free coordinate system, such as
Kruskal-Szekeres, Eddington-Finkelstein, Novikov, Lemaitre\ or other systems
[1-2].

The coordinate system (2.1), though ill-defined on the horizon, may be
applied inside the horizon (see e.g. [6-7]). The interior of a BH, $r<r_{S}$
possesses, apart from some well-known, some not so well-known, properties
too (see [9]). The Killing vector $\partial _{t}$ becomes a spatial one that
results in momentum conservation instead of energy conservation, as obeyed
outside BH (see below). This is accompanied by the interchange of the roles
of the coordinates: $t$ and $r$ play the role of the spatial- and
temporal-like coordinates, respectively.

The interesting feature of the interior of a Schwarzschild BH is that it may
be regarded as a unique spacetime, a cosmological anisotropic model called a
"T-sphere" model or simply T-model [8]. It is described by the line element
(2.1) for $r<r_{S}$ but now expressed in terms of $T\left( =-r\right) $
(temporal) and $z$ (spatial) coordinates instead of $r$ and $t$,
coordinates, respectively

\begin{equation}
ds_{-}^{2}=g_{T}dT^{2}-g_{z}dz^{2}-g_{2}\left( T\right) \left( d\theta
^{2}+\sin ^{2}\theta d\varphi ^{2}\right) ,  \tag{2.4}
\end{equation}%
where, $T\in \left\langle -r_{S},0\right\rangle ,$ $z\in \left( -\infty
,+\infty \right) ,$ $g_{T}=\left( \frac{r_{S}}{T}-1\right) ^{-1}=g_{z}^{-1}$%
. At each instant of $T_{0}$ the spatial slice is a hypercylinder $\mathbf{R}%
^{1}\times S^{2}$, longitudinally expanding and transversally, a two-sphere
of radius $\left\vert T_{0}\right\vert ,$ contracting (see e.g. [6]). Along
the cylinder axis $z$ the system is homogeneous and that represents the
momentum $z$-component conservation.

Phenomena of a classical nature have been considered in the T-model both
within a more traditional approach (see e.g.[10-13]) as well as from other
specific perspectives (see [9], [23-25]). Here we will consider a special
quantum phenomenon, namely the problem of dynamics of the quantized scalar
field in the case T-model will be introduced and briefly discussed within a
unitary approach.

\section{Scalar free field in a T-model}

A scalar free field $\Phi $ in a space-time $M$ with a metric $g_{\mu \nu }$
is described in terms of Lagrangian density $\mathcal{L}$: 
\begin{equation}
\mathcal{L}=\frac{1}{2}\sqrt{-g}g^{\mu \nu }\partial _{\mu }\Phi \partial
_{\nu }\Phi -\left( \mu ^{2}+\xi R\right) \Phi ^{2},  \tag{3.1}
\end{equation}%
\textit{where }$-g=\det \left[ g_{\alpha \beta }\right] ,$ the parameter $%
\mu $ can be interpreted as the mass only in asymptomatically flat
space-time, $R$ is the scalar curvature of $M$ and $\xi $ is the field
coupling to the spacetime curvature.

In the case of the spacetime (2.4) the coupling with gravitational field
vanishes (as $R=0$) and the action of the scalar free field (3.1) takes the
form

\begin{equation}
S=\frac{1}{2}\int dT\int\limits_{\mathbf{\Sigma }}dzd\Omega T^{2}\left[ 
\frac{1}{g_{T}}\left( \partial _{T}\Phi \right) ^{2}-\frac{1}{g_{z}}\left(
\partial _{z}\Phi \right) ^{2}+\frac{1}{T^{2}}\Phi \Delta _{S^{2}}\Phi -\mu
^{2}\Phi ^{2}\right] ,  \tag{3.2}
\end{equation}%
where $\Sigma =\mathbf{R}^{1}\times S^{2},$ $d\Omega =\sin \theta d\varphi
d\theta $ and we have integrated by parts in the sector $S^{2}$ which
resulted in the Laplace operator $\Delta _{S^{2}}$ on $S^{2}$:%
\begin{equation}
\Delta _{S^{2}}\Phi =\frac{1}{\sin \theta }\frac{\partial }{\partial \theta }%
\left( \sin \theta \frac{\partial \Phi }{\partial \theta }\right) +\frac{1}{%
\sin ^{2}\theta }\frac{\partial ^{2}\Phi }{\partial \varphi ^{2}}.  \tag{3.3}
\end{equation}%
The Klein-Gordon (or Euler-Lagrange) equation 
\begin{equation}
\frac{1}{\sqrt{-g}}\partial _{\mu }\left( \sqrt{-g}g^{\mu \nu }\partial
_{\nu }\Phi \right) +\mu ^{2}\Phi =0,  \tag{3.4}
\end{equation}%
takes in this case the following form:%
\begin{equation}
\partial _{T}\left( T^{2}g_{z}\partial _{T}\Phi \right) -\frac{T^{2}}{g_{z}}%
\partial _{z}^{2}\Phi -\Delta _{S^{2}}\Phi +\mu ^{2}T^{2}\Phi =0,  \tag{3.5}
\end{equation}%
Taking the field $\Phi $ in the form of a product:%
\begin{equation}
\Phi \left( T,z,\theta ,\phi \right) =R\left( T\right) u\left( z\right)
Y\left( \theta ,\phi \right) .  \tag{3.6}
\end{equation}%
it follows that the wave equation ( 3.5) separates into the following
equations:%
\begin{equation}
\Delta _{S^{2}}Y=-l\left( l+1\right) Y,  \tag{3.7}
\end{equation}

\begin{equation}
\frac{d^{2}u_{\varepsilon }}{dz^{2}}=-\varepsilon ^{2}u_{\varepsilon }, 
\tag{3.8}
\end{equation}

\begin{equation}
\frac{d}{dT}\left( T^{2}g_{z}\frac{dR_{\varepsilon l}}{dT}\right)
+T^{2}\left( \frac{\varepsilon ^{2}}{g_{z}}+\mu ^{2}+\frac{l\left(
l+1\right) }{T^{2}}\right) R_{\varepsilon l}=0,  \tag{3.9}
\end{equation}%
where $\varepsilon $ is a (separation) constant. The solution of Eq.(3.7)%
\textit{\ }is given by the spherical harmonics $Y_{lm}\left( \theta ,\phi
\right) ,$ 
\begin{equation}
\int\limits_{S^{2}}d\Omega Y_{lm}\left( \theta ,\varphi \right) Y_{l^{\prime
}m^{\prime }}^{\ast }\left( \theta ,\varphi \right) =\delta _{ll^{\prime
}}\delta _{mm^{\prime }},  \tag{3.10}
\end{equation}

\begin{equation}
\int\limits_{S^{2}}d\Omega Y_{lm}\left( \theta ,\varphi \right) Y_{l^{\prime
}-m^{\prime }}\left( \theta ,\varphi \right) =\delta _{ll^{\prime }}\delta
_{m,-m^{\prime }}  \tag{3.11}
\end{equation}%
where $m=-l,-\left( l-1\right) ,...0....,l$. The solution of equation (3.8)
is 
\begin{equation}
u\left( z\right) =e^{\pm i\varepsilon z}.  \tag{3.12}
\end{equation}%
One can decompose the field $\Phi $ into the complete system of functions on 
$\mathbf{R}^{1}$ and $S^{2}$. Thus, the real field $\Phi =\Phi ^{\ast }$ is
represented as:%
\begin{equation}
\Phi \left( T,z,\theta ,\varphi \right) =\sum\limits_{\varepsilon ,l,m}\left[
R_{\varepsilon l}\left( T\right) e^{i\varepsilon z}Y_{lm}\left( \theta
,\varphi \right) A_{\varepsilon lm}+R_{\varepsilon l}^{\ast }\left( T\right)
e^{-i\varepsilon z}Y_{lm}^{\ast }\left( \theta ,\varphi \right)
A_{\varepsilon lm}^{\ast }\right] ,  \tag{3.13}
\end{equation}%
where $R_{\varepsilon l}\left( T\right) $ are the functions of the temporal
variable $T$ satisfying second order differential equation (3.9) and $%
A_{\varepsilon lm}$ are Fourier-like coefficients.

The scalar product $\left( \cdot ,\cdot \right) $ (Klein-Gordon) is in
general defined as :%
\begin{equation}
\left( \Phi ,\Psi \right) =i\int\limits_{\Sigma _{t}}\left( \Phi ^{\ast
}\partial _{\mu }\Psi -\Psi \partial _{\mu }\Phi ^{\ast }\right) n^{\mu
}dvol\left( \Sigma _{t}\right) ,  \tag{3.14}
\end{equation}%
where $n=n^{\mu }\partial _{\mu }$ denotes the unit time-like vector field
orthogonal to a space-like hypersurface (slice) $\Sigma _{t}$ and $\Phi
,\Psi $ are the solutions of the Klein-Gordon equation.\ In this case $%
\Sigma _{t}\simeq A\times S^{2}$ and the scalar product takes the form (see
[17], [26]):%
\begin{equation}
\left( \Phi ,\Psi \right) =iT^{2}g_{z}\int\limits_{S^{2}}\sin \theta d\theta
d\phi \int\limits_{A}\left( \Phi ^{\ast }\partial _{T}\Psi -\Psi \partial
_{T}\Phi ^{\ast }\right) dz.  \tag{3.15}
\end{equation}%
There is the following normalization condition

\begin{equation}
A_{\varepsilon lm}=\left( R_{\varepsilon l}\left( T\right) e^{i\varepsilon
z}Y_{lm}\left( \theta ,\varphi \right) ,\Phi \right)  \tag{3.16}
\end{equation}%
where $\Phi $ is given by (3.6), which is equivalent to the claim of the
canonical commutation relations (see also below).

After some (lengthy but simple) algebra one finds that condition (3.16) is
satisfied iff%
\begin{equation}
T^{2}g_{z}\left[ R_{\varepsilon l}^{\ast }\overset{\cdot }{R}_{\varepsilon
l}-\overset{\cdot }{R}_{\varepsilon l}^{\ast }R_{\varepsilon l}\right] =-i, 
\tag{3.17}
\end{equation}

\begin{equation}
R_{\varepsilon l}^{\ast }\overset{\cdot }{R}_{-\varepsilon l}^{\ast
}-R_{-\varepsilon l}^{\ast }\overset{\cdot }{R}_{\varepsilon l}^{\ast }=0. 
\tag{3.18}
\end{equation}

The condition (3.17) is derived from the differential equation (3.9). First,
one writes Eq.(3.9) for the complex conjugated function $R_{\varepsilon
l}^{\ast }$; then one multiplies it by $R_{\varepsilon l}$ and Eq.(3.9) by $%
R_{\varepsilon l}^{\ast }$; finally one subtracts the former from the latter
obtaining%
\begin{equation}
d_{T}\left( T^{2}g_{z}\left[ R_{\varepsilon l}^{\ast }\overset{\cdot }{R}%
_{\varepsilon l}-\overset{\cdot }{R}_{\varepsilon l}^{\ast }R_{\varepsilon l}%
\right] \right) =0.  \tag{3.19}
\end{equation}%
Therefore, (3.17) turns out to be a normalization condition for $%
R_{\varepsilon l}$ i.e. the Wronskian in this case, as it should be. On the
other hand Eq. (3.18) is just an equivalence.

\subsection{Quantization}

Quantization of the field (3.1-2) is performed in a canonical way. Namely,
one introduces the momentum field as the field canonically conjugated to $%
\Phi \left( T,z,\theta ,\varphi \right) ,$i.e.

\begin{equation}
\pi =\frac{\partial \mathcal{L}}{\partial \left( \partial _{T}\Phi \right) }=%
\frac{T^{2}}{g_{T}}\partial _{T}\Phi .  \tag{3.20}
\end{equation}%
Then one imposes canonical commutation relations 
\begin{eqnarray}
\left[ \widehat{\Phi }\left( t,\mathbf{x}\right) ,\widehat{\pi }\left( t,%
\mathbf{y}\right) \right] &=&i\delta \left( \mathbf{x,y}\right) , 
\TCItag{3.21} \\
\left[ \widehat{\Phi }\left( t,\mathbf{x}\right) ,\widehat{\Phi }\left( t,%
\mathbf{y}\right) \right] &=&\left[ \widehat{\pi }\left( t,\mathbf{x}\right)
,\widehat{\pi }\left( t,\mathbf{y}\right) \right] =0,  \notag
\end{eqnarray}%
where $\mathbf{x,y}\in \Sigma _{t}$. In our case the slice $\Sigma _{t}$ has
the topology of the product space of the set $A\subset \mathbf{R}^{1}$ and
the two-dimensional sphere $S^{2}$. The momentum field given in its Fourier
decomposed form is:%
\begin{equation}
\widehat{\pi }\left( t,r,\theta ,\phi \right) =\frac{T^{2}}{g_{T}}%
\sum\limits_{\varepsilon ,l,m}\left[ \widehat{A}_{\varepsilon lm}\overset{%
\cdot }{R}_{\varepsilon l}\left( T\right) e^{i\varepsilon z}Y_{lm}\left(
\theta ,\phi \right) +\widehat{A}_{\varepsilon lm}^{\dag }\overset{\cdot }{R}%
_{\varepsilon l}^{\ast }\left( T\right) e^{-i\varepsilon z}Y_{lm}^{\ast
}\left( \theta ,\phi \right) \right]  \tag{3.22}
\end{equation}

The canonical commutation relations Eqs. (3.21) turn out to be satisfied
under the following conditions:

a) $\widehat{A}_{\varepsilon lm}$, $\widehat{A}_{\varepsilon lm}^{\dag },$
are the annihilation and creation operators, respectively, i.e. the only
nonvanishing commutator is%
\begin{equation}
\left[ \widehat{A}_{\varepsilon lm},\widehat{A}_{\varepsilon ^{\prime
}l^{\prime }m^{\prime }}^{\dag }\right] =\delta _{\varepsilon \varepsilon
^{\prime }}\delta _{ll^{\prime }}\delta _{mm^{\prime }}  \tag{3.22}
\end{equation}

b) the Wronskian (3.17) must hold.

\section{Hamiltonian of the scalar field in a T-model}

The Hamiltonian of the field described by the Lagrangian density $\mathcal{L}
$\ is determined as an integral over the spatial part $\mathbf{\Sigma }$\ of
the spacetime

\begin{equation}
H=\int\limits_{\mathbf{\Sigma }}d^{3}x\left[ \pi \partial _{T}\Phi -\mathcal{%
L}\right] ,  \tag{4.1}
\end{equation}%
and this expression is equivalent to the (integrated) $T_{TT}$ element of
the stress-energy tensor. Applying formula (4.1) for the case (2.4) and
(3.1) one obtains

\begin{equation}
H=\frac{1}{2}\int\limits_{\mathbf{\Sigma }}dzd\theta d\varphi T^{2}\sin
\theta \left[ \frac{1}{g_{T}}\left( \partial _{T}\Phi \right) ^{2}+\frac{1}{%
g_{z}}\left( \partial _{z}\Phi \right) ^{2}-\Phi \Delta _{S^{2}}\Phi +\mu
^{2}\Phi ^{2}\right]  \tag{4.2}
\end{equation}%
Using the Fourier decomposition of the quantized field and momentum (see
Eqs. (3.13), (3.22)) one finds \ the Hamiltonian of the quantized scalar
field as expressed in terms of annihilation and creation operators:%
\begin{equation}
H=\frac{1}{2}\sum\limits_{\varkappa }\left[ \omega _{\varkappa }\widehat{A}%
_{\varkappa }\widehat{A}_{\varkappa }^{\dag }+\gamma _{\varkappa \varkappa
^{\prime }}\widehat{A}_{\varkappa }\widehat{A}_{\varkappa ^{\prime }}+\left(
c.c\right) \right]  \tag{4.3}
\end{equation}%
where indices $\varkappa ,\varkappa ^{\prime }$ correspond to the
appropriate three letter sets $\varepsilon lm.$ The parameters $\omega
_{\varkappa },\gamma _{\varkappa \varkappa ^{\prime }}$\ are given as

\begin{equation}
\gamma _{\varepsilon lm/\varepsilon ^{\prime }lm^{\prime }}=\left[ T^{2}g_{z}%
\overset{\cdot }{R}_{\varepsilon l}\overset{\cdot }{R}_{-\varepsilon
l}+T^{2}\left\{ \frac{\varepsilon ^{2}}{g_{z}}+\frac{l\left( l+1\right) }{%
T^{2}}+\mu ^{2}\right\} R_{\varepsilon l}\left( T\right) R_{-\varepsilon
l}\left( T\right) \right] \delta _{\varepsilon ,-\varepsilon ^{\prime
}}\delta _{m,-m^{\prime }}  \tag{4.4}
\end{equation}%
\bigskip

\begin{equation}
\omega _{\varepsilon lm}=\left[ T^{2}g_{z}\overset{\cdot }{R}_{\varepsilon l}%
\overset{\cdot }{R}_{\varepsilon l}^{\ast }+T^{2}\left\{ \frac{\varepsilon
^{2}}{g_{z}}+\frac{l\left( l+1\right) }{T^{2}}+\mu ^{2}\right\}
R_{\varepsilon l}\left( T\right) R_{\varepsilon l}^{\ast }\left( T\right) %
\right] .  \tag{4.5}
\end{equation}%
Therefore, the Hamiltonian of the scalar field in the T-model, i.e.
anisotropic cosmological model representing interior of the Schwarzschild
BH, turns out to be

\begin{equation}
H=\frac{1}{2}\sum\limits_{\varepsilon lm}\left[ \omega _{\varepsilon
lm}\left( \widehat{A}_{\varepsilon lm}\widehat{A}_{\varepsilon lm}^{\dag }+%
\widehat{A}_{\varepsilon lm}^{\dag }\widehat{A}_{\varepsilon lm}\right)
+\gamma _{\varepsilon lm/-\varepsilon l-m}\widehat{A}_{\varepsilon lm}%
\widehat{A}_{-\varepsilon l-m}+\gamma _{\varepsilon lm/-\varepsilon
l-m}^{\ast }\widehat{A}_{\varepsilon lm}^{\dag }\widehat{A}_{-\varepsilon
l-m}^{\dag }\right] .  \tag{4.6}
\end{equation}%
representing the set of interacting, time-dependent harmonic oscillators.

On this basis one can study the dynamics of the quantized scalar field. The
evolution of the system is described by the Heisenberg equation of motion
for the operators $\widehat{A}_{\varepsilon lm}$%
\begin{equation}
i\frac{d}{dt}\widehat{A}_{\varepsilon lm}=\left[ \widehat{A}_{\varepsilon
lm},\widehat{H}\right] =\omega _{\varepsilon lm}\left( t\right) \widehat{A}%
_{\varepsilon lm}\left( t\right) +\gamma _{\varepsilon lm}^{\ast }\left(
t\right) \widehat{A}_{-\varepsilon l-m}^{\dag }\left( t\right)  \tag{4.7}
\end{equation}%
where, $\gamma _{\varepsilon lm/-\varepsilon l-m}\equiv \gamma _{\varepsilon
lm}.$ One can search for the solutions of the above equations by using the
following ansatz:

\begin{equation}
\widehat{A}_{\varepsilon lm}\left( t\right) =\alpha _{\varepsilon lm}\left(
t\right) \widehat{A}_{\varepsilon lm}+\beta _{\varepsilon lm}\left( t\right) 
\widehat{A}_{-\varepsilon l-m}^{\dag },  \tag{4.8}
\end{equation}%
where $\alpha _{\varepsilon lm}\left( t\right) $ and $\beta _{\varepsilon
lm}\left( t\right) $ are some unknown complex functions and $\widehat{A}%
_{\varepsilon lm}$ and $\widehat{A}_{-\varepsilon l-m}^{\dag }$ are time
independent operators. By definition the relation (4.8) preserves the
commutation relations (3.22), hence it turns out to be the Bogolyubov
transformation,%
\begin{equation}
\left\vert \alpha _{\varepsilon lm}\left( t\right) \right\vert
^{2}-\left\vert \beta _{\varepsilon lm}\left( t\right) \right\vert ^{2}=1. 
\tag{4.9}
\end{equation}%
Then, the Heisenberg equations (4.7) are converted into differential
equations for the Bogolyubov coefficients

\begin{equation}
i\frac{d}{dt}\alpha _{\varepsilon lm}\left( t\right) =\omega _{\varepsilon
lm}\left( t\right) \alpha _{\varepsilon lm}\left( t\right) +\gamma
_{\varepsilon lm}^{\ast }\left( t\right) \beta _{\varepsilon lm}^{\ast
}\left( t\right) ,  \tag{4.10}
\end{equation}

\begin{equation}
i\frac{d}{dt}\beta _{\varepsilon lm}\left( t\right) =\omega _{\varepsilon
lm}\left( t\right) \beta _{\varepsilon lm}\left( t\right) +\gamma
_{\varepsilon lm}^{\ast }\left( t\right) \alpha _{\varepsilon lm}^{\ast
}\left( t\right) .  \tag{4.11}
\end{equation}%
In general, one can't expect exact solutions of the equations (4.10-11) and
approximate schemes would therefore be proposed. Our forthcoming paper will
be devoted to the comprehenssive discussion of this problem.

\section{Discussion}

\ Considering the interior of a Schwarzschild BH as a unique spacetime, an
anisotropic cosmological model, we have performed the quantization of the
free (noninteracting) scalar field by imposing the canonical commutation
relations. One decomposes the field and momentum in terms of the complete
set of solutions of the Klein-Gordon (or in fact Euler-Lagrange equations)
with the coefficients of expansion being annihilation and creation
operators. This procedure leads to the Hamiltonian of the quantized scalar
field taking the form of the set of harmonic, time-dependent oscillators
coupled in a special way: there are terms in the Hamiltonian corresponding
to creation, $\gamma _{\varepsilon lm}\widehat{A}_{\varepsilon lm}\widehat{A}%
_{-\varepsilon l-m}$ and annihilation $\gamma _{\varepsilon lm}^{\ast }%
\widehat{A}_{\varepsilon lm}^{\dag }\widehat{A}_{-\varepsilon l-m}^{\dag }$
particles in pairs.

Such a picture, peculiar at first sight, appears to have a deeper sense. The
spacetime considered is a dynamic one - there is no energy conservation
there, hence the Hamiltonian contains terms representing spontaneous
creation and annihilation pairs of particles. Homogeneity of the spacetime
along the $z$-direction results in the presence of a spatial-like Killing
vector representing, $z$-momentum-component conservation. Hamiltonian (4.6)
reflects this symmetry property: pairs of the particles with opposite $z$%
-component momenta may be created, $\widehat{A}_{\varepsilon lm}^{\dag }%
\widehat{A}_{-\varepsilon l-m}^{\dag }$ and annihilated $\widehat{A}%
_{\varepsilon lm}\widehat{A}_{-\varepsilon l-m}$; the Hamiltonian of the
system also obeys rotational invariance.\ 

The conservation of the $z$-momentum component in the terms represented by $%
\gamma _{\varepsilon lm}$\ and $\gamma _{\varepsilon lm}^{\ast }$\ in the
Hamiltonian is an analogue of energy conservation outside the BH, i.e. the
particles in a pair carry positive/negative energy; the one with negative
energy cannot survive outside the BH but only within the horizon of the BH.

There is a more or less obvious interpretation of the $\beta _{\varepsilon
lm}\left( t\right) $\ coefficient of the Bogolyubov's transformation (4.8):
it is proportional to the number of the particles created during the
evolution of the system,%
\begin{equation}
\left\langle 0\left( t\right) \right\vert \widehat{A}_{\varepsilon lm}^{\dag
}\widehat{A}_{\varepsilon lm}\left\vert 0\left( t\right) \right\rangle
=\left\langle 0\right\vert \widehat{A}_{\varepsilon lm}^{\dag }\left(
t\right) \widehat{A}_{\varepsilon lm}\left( t\right) \left\vert
0\right\rangle =\left\vert \beta _{\varepsilon lm}\left( t\right)
\right\vert ^{2},  \tag{5.1}
\end{equation}%
where $\left\vert 0\right\rangle $\ is the vacuum state for fixed time $t=0$%
\ and annihilation operators $\widehat{A}_{\varepsilon lm}$\ while $%
\left\vert 0\left( t\right) \right\rangle $\ is the vacuum state for later
time $t$\ and annihilation operators $\widehat{A}_{\varepsilon lm}\left(
t\right) .$ Due to the violent dynamics of the background spacetime, one may
expect the dynamics of the creation and annihilation of the (pairs of)
particles to be violent, and conventional adiabatic-like approaches (see
e.g. [17-18]) could hardly be regarded as a working scheme. Therefore,
attempts to find an approximate solution within a treatment here proposed
that might be called a "unitary approach" as based on a unitarity of the
evolution of the system, will be discussed in our following paper.

An interesting aspect of the dynamics of the model (3.1) will be however
briefly discussed here. That is the question of the possible instability of
the system of interacting harmonic oscillators (4.6) (see [27-28]). The
oscillators interact in pairs, $\left( \varepsilon lm\right) /\left(
-\varepsilon l-m\right) $ and one can consider diagonalization (at an
arbitrary instant $T^{\prime }$) of the Hamiltonian corresponding to such a
subsystem. Then the frequency in such a diagonalized case is given as:%
\begin{equation}
\Omega _{\varepsilon lm}^{2}=\omega _{\varepsilon lm}^{2}-\left\vert \gamma
_{\varepsilon lm/-\varepsilon l-m}\right\vert ^{2}.  \tag{5.2}
\end{equation}

This expression should be positive, otherwise the system is unstable (see
[27]) (this problem will be discussed in detail in our following paper) -
this would be named a "gravitationally driven instability". One can check
that in this case, Eqs. (4.4-5) the right hand side of Eq.(5.2)\ 
\begin{equation}
\Omega _{\varepsilon lm}^{2}=\frac{1}{g_{z}}\left[ \frac{\varepsilon ^{2}}{%
g_{z}}+\frac{l\left( l+1\right) }{T^{2}}+\mu ^{2}\right]   \tag{5.3}
\end{equation}%
is positive: there is no gravitational instability in the scalar field
quantized in Doran et al. [7] spacetime. An interesting issue is that, apart
from the possible instability of type (5.2), that might be referred to as "a
restoring force instability" there is also another possible instability,
namely "a friction driven instability" but the problem of its origin and
character will be discussed elsewhere.

\section{Appendix}

Let us briefly analyze the form of the temporal part of Klein-Gordon
equation in this case, i.e. Eq. (3.9):%
\begin{equation}
\frac{1}{T^{2}}\frac{d}{dT}\left( T^{2}g_{z}\frac{dR}{dT}\right) +\left( 
\frac{\varepsilon ^{2}}{g_{z}}+\mu ^{2}+\frac{l\left( l+1\right) }{T^{2}}%
\right) R=0,  \tag{A.1}
\end{equation}%
where $g_{z}=\frac{r_{S}}{T}-1,$ and \ lower labels have been omitted here.
Making the substitution, $R=f\eta ,$ one finds%
\begin{equation}
\frac{1}{T^{2}}\frac{d}{dT}\left( T^{2}g_{z}\frac{dR}{dT}\right) =\frac{1}{%
T^{2}}\left[ \left( r_{S}-2T\right) \left( f^{\prime }\eta +f\eta ^{\prime
}\right) +\left( r_{S}T-T^{2}\right) \left( f^{\prime \prime }\eta +f\eta
^{\prime \prime }+2f^{\prime }\eta ^{\prime }\right) \right]  \tag{A.2}
\end{equation}%
and prime means differentiation with respect to $T$. Claiming%
\begin{equation}
\left( r_{S}-2T\right) f+2\left( r_{S}T-T^{2}\right) f^{\prime }=0, 
\tag{A.3}
\end{equation}%
one gets $R\left( T\right) $ in the form

\begin{equation}
R=\frac{\eta }{\sqrt{T\left( r_{S}-T\right) }},  \tag{A.4}
\end{equation}%
and $\eta \left( T\right) $ satisfies the following confluent Heun equation%
\begin{equation}
\left[ \frac{d^{2}}{dT^{2}}+\nu ^{2}\left( T\right) \right] \eta =0, 
\tag{A.5}
\end{equation}%
where%
\begin{equation}
\nu ^{2}\left( T\right) =A+\frac{B}{T}+\frac{C}{\left( r_{S}-T\right) }+%
\frac{D}{T^{2}}+\frac{E}{\left( r_{S}-T\right) ^{2}},  \tag{A.6}
\end{equation}%
and the five coefficients $A,...,E$ are equal to: 
\begin{gather*}
A=\left( \varepsilon ^{2}-\mu ^{2}\right) ,\text{ \ \ }B=\frac{1}{2r_{S}}%
\left( 2l\left( l+1\right) +1\right) , \\
C=r_{S}\left( \mu ^{2}+2\varepsilon ^{2}\right) +B,\text{ \ \ }D=\frac{1}{4},
\\
E=D-2\left( 1+r_{S}^{2}\varepsilon ^{2}\right) .
\end{gather*}

\section{References}

[1] V.P. Frolov, V.P. and I.D. Novikov, \textit{Black Hole Physics: Basic
Concepts and New Developments}; Kluwer \qquad Academic: Dordrecht, The
Netherlands, 1998

[2] J.B. Hartle, \textit{Gravitation}, Addison Wesley, 2003

[3] M. Ba\'{n}ados, J. Silk, and S.M. West, \textit{Kerr Black Holes as
Particle Accelerators to Arbitrarily High Energy}, Phys. Rev. Lett. 103,
111102 (2009); arXiv:0909.0169

[4] T. Harada and M. Kimura, \textit{Black holes as particle accelerators: A
brief review}, Classical Quantum Gravity 31, 243001 (2014)

[5] O.B. Zaslavskii \textit{The Ba\'{n}ados-Silk-West effect with immovable
particles near static black holes and its rotational counterpart};
arXiv:2207.03213v2 [gr-qc] (2022), https://doi.org/10.48550/arXiv.2207.03213.

[6] P. Gusin, A. T. Augousti, F. Formalik, and A. Radosz, \textit{The
(A)symmetry between the Exterior and Interior of a Schwarzschild Black Hole}%
, Symmetry (2018) 10, 366

[7] R. Doran, F. S. Lobo, and P. Crawford, \textit{Interior of a
Schwarzschild black hole revisited}, Foundations of Physics, vol. 38, no. 2,
pp. 160 (2008)

[8] V.A. Ruban, Spherically Symmetric T-Models in the General Theory of
Relativity, Gen. Rel. and Grav., 33, No. 2, 2001

[9] A.J.S. Hamilton and G. Polhemus, \textit{Stereoscopic visualization in
curved spacetime: Seeing deep inside a black hole}, New J. Phys. (2010), 12,
123027--123052

[10] A. Radosz, P. Gusin, A. T. Augousti and F. Formalik, \textit{Inside
spherically symmetric black holes or how a uniformly accelerated particle
may slow down} Eur. Phys. J. C 2019, 79, 876.
https://doi.org/10.1140/epjc/s10052-019-7372-5

[11] A.V. Toporensky and O.B. Zaslavskii, \textit{Zero-momentum trajectories
inside a black hole and high energy particle collisions}, J. Cosmol. Astr.
Phys. 2019(12):063-063.

[12] A. T. Augousti, P. Gusin, B. Ku\'{s}mierz, J. Masajada and A. Radosz, 
\textit{On the speed of a test particle inside the Schwarzschild event
horizon and other kinds of black holes}. Gen. Relativ. Gravit. 2018, 50,
131, doi:10.1007/s10714-018-2445-6.

[13] A. V. Toporensky and O. B. Zaslavskii, \textit{Redshift of a photon
emitted along the black hole horizon}," E.

Phys. J. C, vol. 77, no. 3, p. 179, 201

[14] D. V. Gal'tsov and E. E. Donets, \textit{Power--law mass inflation in
Einstein--Yang--Mills--Higgs black holes, }arXiv:gr-qc/9706067v1 22 Jun 1997

[15] E.E. Donets, D. V. Gal'tsov and M.Y. Zotov, \textit{Internal Structure
of Einstein--Yang--Mills Black Holes, }https://arxiv.org/pdf/gr-qc/9612067

[16] U. A. Yajnik and K. Narayan, \textit{CPT invariance and canonical
quantization inside the Schwarzschild black hole, }Class. Quantum Grav. 15
1315, 1998

[17] E. Parker, D. J. Toms, \textit{Quantum Field Theory In Curved
Spacetime, Quantized Fields and Gravity}, 2009

[18] S. Habib, C. Molina-Paris, E. Mottola, \textit{Energy-Momentum Tensor
of Particles Created in an Expanding Universe}, Phys.Rev. D61 (2000) 024010,
[arXiv:gr-qc/9906120]

[19] G. Tsoupros, \textit{Conformal Scalar Propagation inside the
Schwarzschild Black Hole, }Gen.Rel.Grav. 44 (2012) 309-351

[20] N. Oshita, \textit{Resolution to the firewall paradox: The black hole
information paradox and highly squeezed interior quantum fluctuations, }%
Class.Quant.Grav. 34, 19, 195002, 2017

[21] C. R. Almeida and D. C. Rodrigues, \textit{Quantization of a black-hole
gravity: geometrodynamics and the quantum, }Class.Quant.Grav. 40, 3, 035004,
2023

[22] S. B. Giddings and J. Perkins, \textit{Quantum evolution of the Hawking
state for black holes}, \qquad arXiv.2204.1312 [hep-th]

[23] Christodoulou, M.; Rovelli, C. \textit{How big is a black hole?} Phys.
Rev. D 2015, 91, 064046

[24] P. Gusin, A. Radosz, \textit{The volume of the black holes - The
constant curvature slicing of the spherically symmetric spacetime}, Mod.
Phys. Lett. A 32, 1750115 (2017)

[25] Zaslavskii, O.B. \textit{Schwarzschild Black Hole as Accelerator of
Accelerated Particles} JETP Lett. 2020, 111, doi:10.1134/S0021364020050033

[26] N. D. Birrel, P. C. W. Davies, \textit{Quantum Fields in Curved Space},
Cambridge 1984.

[27] Rajeev, K., Chakraborty, S. \& Padmanabhan, T. \textit{Inverting a
normal harmonic oscillator: physical interpretation and applications}, Gen
Relativ Gravit 50, 116 (2018). https://doi.org/10.1007/s10714-018-2438-5

[28] Ya. B. Zel'dovich and A. A. Starobinsky, \textit{Particle production
and vacuum polarization in an anisotropic gravitational field}, Sov. Phys.
(JETP) 34, 1159 (1972)

\end{document}